\documentclass[conference]{IEEEtran}

\usepackage{amsmath,amssymb,amsfonts}
\usepackage{algorithm}
\usepackage{courier}
\usepackage{graphicx}
\usepackage{textcomp}
\usepackage{xcolor}
\usepackage[utf8]{inputenc}
\usepackage{listings}
\usepackage{float}
\usepackage{subfigure}
\usepackage[hyphens]{url}
\usepackage{multirow}
\usepackage{commath}
\usepackage{algpseudocode}
\usepackage{etoolbox}
\usepackage{varwidth}
\usepackage{verbatim}
\usepackage{tikz}
\usetikzlibrary{shapes,arrows}
\usepackage{colortbl}
\usepackage{balance}

\ifCLASSINFOpdf
\else
\fi

\begin{document}
% paper title

\title{Comparison of Vectorization Capabilities of Different Compilers for X86 and ARM CPUs}

% author names and affiliations
% use a multiple-column layout for up to three different
% affiliations
%

\author{
    \IEEEauthorblockN{Nazmus Sakib\IEEEauthorrefmark{1}, 
    Tarun Prabhu\IEEEauthorrefmark{2}, 
    Nandakishore Santhi\IEEEauthorrefmark{2},
    John Shalf\IEEEauthorrefmark{3},
    Abdel-Hameed A. Badawy\IEEEauthorrefmark{1}}

    \IEEEauthorblockA{\IEEEauthorrefmark{1}Klipsch School of ECE, New Mexico State University, Las Cruces, NM 88003, USA \\
    %Email: \{nsakib6, badawy\}@nmsu.edu
    }
    \IEEEauthorblockA{\IEEEauthorrefmark{2}Los Alamos National Laboratory, Los Alamos, NM 87545, USA \\
    %Email: \{tarun, nsanthi\}@lanl.gov
    }
    \IEEEauthorblockA{\IEEEauthorrefmark{3}Lawrence Berkeley National Laboratory, Berkeley, CA 94720, USA \\
    %Email: jshalf@lbl.gov
    }
    \IEEEauthorblockA{\IEEEauthorrefmark{1}\{nsakib6, badawy\}@nmsu.edu, \IEEEauthorrefmark{2}\{tarun, nsanthi\}@lanl.gov, \IEEEauthorrefmark{3}jshalf@lbl.gov}    
}

% use for special paper notices
%\IEEEspecialpapernotice{(Invited Paper)}
\newcommand{\code}[1]{{\footnotesize\texttt{#1}}}
\newcommand{\keyword}[1]{{\footnotesize\texttt{\textbf{#1}}}}
\newcommand{\asm}[1]{{\footnotesize\texttt{\textit{#1}}}}
\newcommand{\reg}[1]{{\footnotesize\texttt{\textit{#1}}}}

%% Global settings for listings.
\lstdefinestyle{asm}{
   language=[x86masm]Assembler,
   xleftmargin=12pt,
   numbers=left,
   numbersep=4pt,
   basicstyle=\footnotesize\ttfamily,
   morekeywords={vmovaps, vmovups,vcmpltps,vfmadd213ps,vmovss,vcomiss,vfmadd132ss,vaddps,vaddss,vfmadd231ss,nopw,ldr,str,fmov,b.ne,stur,ldur,ld1w,st1w,fmad,prfm,fmadd,eor,cmpeq,sel,vgatherdps,index,vpmullq,kxnorw,vxorps,vgatherqps,vscatterqps,vpaddq,vmulps,vextractf64x4,vextractf128,vshufpd,vmovshdup,vmulss,fadda,whilelo,b.mi,ldpsw,ldp,scvtf}
}

\lstdefinestyle{c}{
   language=C,
   xleftmargin=12pt,
   numbers=left,
   tabsize=2,
   stepnumber=1,
   numbersep=4pt,
   basicstyle=\footnotesize\ttfamily
}

% make the title area
\maketitle
\thispagestyle{plain}
% As a general rule, do not put math, special symbols or citations
% in the abstract
\begin{abstract}
Most modern processors contain vector units that simultaneously perform the same arithmetic operation over multiple sets of operands. The ability of compilers to automatically vectorize code is critical to effectively using these units. Understanding this capability is important for anyone writing compute-intensive, high-performance, and portable code. We tested the ability of several compilers to vectorize code on x86 and ARM. We used the TSVC2 suite, with modifications that made it more representative of real-world code. On x86, GCC reported $54\%$ of the loops in the suite as having been vectorized, ICX reported $50\%$, and Clang, $46\%$. On ARM, GCC reported $56\%$ of the loops as having been vectorized, ACFL reported 
$54\%$, and Clang, $47\%$. We found that the vectorized code did not always outperform the unvectorized code. In some cases, given two very similar vectorizable loops, a compiler would vectorize one but not the other. We also report cases where a compiler vectorized a loop on only one of the two platforms. Based on our experiments, we cannot definitively say if any one compiler is significantly better than the others at vectorizing code on any given platform.
\end{abstract}

% no keywords

% For peer-reviewed papers, you can put extra information on the cover
% page as needed:
% \ifCLASSOPTIONpeerreview
% \begin{center} \bfseries EDICS Category: 3-BBND \end{center}
% \fi
%
% For peer-review papers, this IEEEtran command inserts a page break, and
% creates the second title. It will be ignored for other modes.
\IEEEpeerreviewmaketitle
\section{Introduction}
Early supercomputers, like the Cray machines, had vector units to take advantage of the data parallelism common to many computationally intensive scientific applications. Such vector units exist on most processors. %and some even have additional support to boost %machine learning applications, which also %benefit from vectorization~\cite{dl}.%

Maximizing the utilization of these vector units requires the appropriate use of vector instructions. However, programming in a high-level language like C with platform-specific vector intrinsics or writing assembly by hand is cumbersome, error-prone, not portable, and unlikely to result in optimal performance unless done by an expert. 

High-quality vectorizing compilers are more likely to produce correct, high-performance code. Some also support generating code for different hardware platforms, allowing a programmer to obtain vectorized code for any platform supported by the compiler from a single code base.

GCC~\cite{urlgcc} and Clang~\cite{urlclang} are widely used open-source compilers that can generate code for X86 and ARM (among other platforms). The Intel oneAPI DPC++/C++ Compiler~\cite{urlicx} and ARM Compiler for Linux (ACFL)~\cite{urlacfl} are proprietary, vendor-provided compilers for X86 and ARM, respectively. In this paper, we compare the ability of these compilers to vectorize on two widely used hardware platforms and compare the performance of the resulting code. We also perform a detailed analysis of the code generated by these compilers in cases where one significantly outperforms the others.
Prior studies~\cite{5575631,6914394} have compared the size and relative performance of the code generated by some of these compilers but not their vectorization abilities. Others~\cite{new:pa, Maleki:evc:pac,last:taco, vid:mu, feng2021evaluation} have studied the compiler's ability to vectorize, but none evaluated the same compiler on different hardware platforms. Pohl~\textit{et al.}~\cite{pohl2020vectorization} have studied the accuracy of speedup prediction by compilers on different platforms, but they did not compare the compilers' ability to vectorize. 

\section{Evaluation}
In this section, we discuss our choice of compilers and hardware platforms.

\begin{table}[b]
    \centering
        %\caption{Compiler and their versions}
        \caption{Compilers and their versions}
        \vspace{-2mm}
        \begin{tabular}{c|c}
            {\textbf { Name }} & {\textbf { Version }}\\
            \cline{1-2}\\[-2.5mm]
            GCC  & {14.1.1}\\
            Clang  & {18.1.8}\\
            ICX  & {2024.0.2}\\
            ACFL  & {22.2}\\
            \cline{1-2}
        \end{tabular}
        %\vspace*{4mm}
    \label{fig:versions}
\end{table}

\subsection{Compilers}
Table~\ref{fig:versions} lists the versions of the compilers used in our experiments. For the vendor-provided compilers, we used the latest versions that were available on our test system. Note that the ACFL 22.2 is based on LLVM 13.0.1, which is older than the Clang version that was used. We expected the two vendor-provided compilers, ICX on x86 and ACFL on ARM, to outperform the open-source compilers, GCC and Clang. 

%
%\begin{figure}[t]
 % \centering
  %\lstinputlisting[style=c]{tsvc_example.c}
  %\vspace{-3mm}
  %\caption{An Example TSVC function}
  %\label{fig:tsvc_example}
%\end{figure}
%
\subsection{Benchmark}
TSVC (Test Suite for Vectorizing Compilers)~\cite{Callahan:tsvc} is a well-known benchmark suite that is used to assess a compiler's ability to vectorize. This suite consists of 151 loop nests containing a variety of control-flow and memory-access patterns such as conditional branches, non-unit strides, reverse array accesses, indirect memory accesses, etc. 
A variant of TSVC is TSVC2~\cite{tsvc2} which utilizes modern C features and prevents function inlining.
%a variant of TSVC with changes to prevent functions from being inlined.% since doing so could affect a compiler's decision to vectorize and the vectorization strategies it employs.%

In TSVC2, each loop nest is contained within a function with exactly one loop nest per function. Since every function contains exactly one loop nest, we use the name of the containing function to refer to a loop nest. 

The loop nests generally perform 32-bit floating point operations on one or more arrays. An outer loop wraps the nest, resulting in redundant computations. This is done to minimize the effect of noise and jitter on the timings and accommodate systems that lack high-resolution timers. %An external function is called at the end of every outer loop iteration. This prevents the compilers from eliminating the loop as dead code.

In TSVC2, the arrays operated on by the loops are global with sizes that are known at compile time. The trip counts of the loop are also compile-time constants. This is not representative of scientific applications where the arrays are usually dynamically allocated and the trip counts are often input-dependent. In order to obtain a more realistic assessment of the compilers i.e. how they performed on real-world code, the code in TSVC2 was modified so the array sizes and the loop trip counts would not be compile-time constants~\cite{tsvc_withArgs}. This was achieved by bundling them together in a \code{struct} which was then passed to the functions. This ensured that the compiler would have to perform more sophisticated analyses to ensure the safety and accurately estimate the profitability of optimizations such as vectorization and loop unrolling. Siso~\textit{et al.}~\cite{siso2019evaluating} demonstrated the effect of withdrawing some compile-time information such as globally known array bounds. Here, we withdraw all global information.
%This eliminates the need for the compiler to carry out expensive aliasing checks or specialize code for unknown trip counts, which are known to impact a compiler's ability to vectorize negatively. % The \code{dummy} and \code{calc\_checksum} functions to prevent the compiler from marking the loop as a dead code and eliminating it altogether. In our experiments, all arrays used in the computation consist of 4-byte floats.% 
%Figure~\ref{fig:tsvc_example} shows one function from the suite.%

\subsection{Hardware}
\begin{table}
    \centering
    \caption{Hardware Specification}
    \vspace{-1mm}
        \begin{tabular}{c|c|c|c}
            {\textbf { Name }} & {\textbf { Architecture }} & {\textbf { Model }} & {\textbf { Vendor }}\\
            \cline{1-4}\\[-2.5mm]
            {\text { Intel }}  & {\text{x86\_64}} & {\text{Xeon(R) Gold 6152}} & {\text {Intel}}\\
            {\text { ARM }}  & {\text{aarch64}} & {\text{A64FX}} & {\text{ Fujitsu}}\\
            \cline{1-4}
        \end{tabular}
        %\vspace*{4mm}
    \label{fig:hw}
\vspace*{-4mm}
\end{table}

Details about the hardware platforms on which we carried out the experiments are provided in Table~\ref{fig:hw}.

\section{Results and Analysis}
\subsection{X86}\label{y}
GCC and Clang were passed \code{\footnotesize-O3 -march=skylake-avx512 -mprefer-vector-width=512} when compiling the test-suite. The last option was replaced with \code{\footnotesize -qopt-zmm-usage=high} on ICX.
The vectorization reports were generated with \code{\footnotesize-fopt-info-all} on GCC, \code{\footnotesize-Rpass=loop-vectorize -Rpass-missed=loop-vectorize} on Clang, and \code{\footnotesize-qopt-report} on ICX.
\begin{figure}[t]
\centering
\begin{tikzpicture}
%% You can adjust the opacity here. For venn diagrams, it is convenient to have a low opacity so that you can see intersections
	\begin{scope} [fill opacity = .4]
%% The draw command knows a lot of shapes. To make a rectangle, you need to specify two diagonal corners. Ensure you always have a semicolon at the end of your draw commands, otherwise latex flips out.
    %\draw (-2,2.5) rectangle (2,-0.5);
    \draw (0,0) rectangle (5,5);
%% Similarly, you can make a circle by specifying the center and then the radius. You can also add a fill color, but if you're printing in black and white you'll probably want to remove that line.
    \draw[fill=green, draw = black] (2.5,3) circle (1);
    \draw[fill=blue, draw = black] (1.8,2.2) circle (1);
    \draw[fill=red, draw = black] (3.2,2.2) circle (1);
    %\draw[fill=red, draw = black] (0,-2) circle (3);
%% We can use the node command to label points. If you put your cursor on "LARGE" or "textbf" a box will drop down with size and text style options.
    %\node at (-2,3.2) {\LARGE\textbf{X}};
    \node at (2.5,4.5) {\color{black} \textbf{ICX}};
    \node at (3.4,4.5) {\color{black}\textbf{(50\%)}};
    \node at (4,1) {\color{black}\textbf{GCC}};
    \node at (4,0.65) {\color{black}\textbf{(54\%)}};
    \node at (1,1) {\color{black}\textbf{Clang}};
    \node at (1,0.65) {\color{black}\textbf{(46\%)}};
    \node at (2.5,2.4) {\textbf{53}}; %all y%
    \node at (2.5,1.8) {\textbf{12}}; %gycyin%
    \node at (1.5,1.8) {\textbf{2}}; %cy%
     \node at (2,2.7) {\textbf{2}}; %gncyiy%
    \node at (2.5,3.5) {\textbf{11}}; %iy%
    \node at (3,2.7) {\textbf{9}}; %gycniy y%
    \node at (3.5,1.8) {\textbf{8}}; %gy%
    \node at (0.8,4.8) {\textbf{Total: 151}};
     \node at (0.75,4.15) {\textbf{None: 54}};
     
    \end{scope}
%% And now you have a venn diagram. Yay!
%\draw[help lines](-5,5) grid (5,-6);    This line can draw the grid lines to help guide you. I use these when I'm writing the code and then delete this line when I publish the pdf.
\end{tikzpicture}
    \caption{Loops vectorized by GCC, ICX, and Clang on x86}
    \label{fig:ven}
\end{figure}
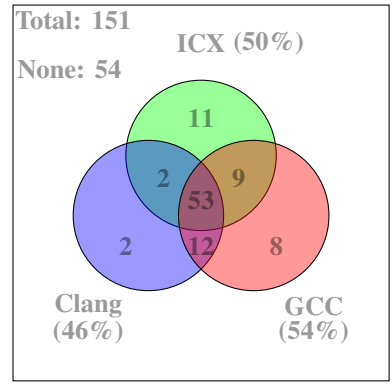

Figure~\ref{fig:ven} shows the number of loops vectorized by GCC, Clang, and ICX. Out of the 151 loops, GCC did not vectorize $46\%$, Clang did not vectorize $54\%$ and ICX did not vectorize $50\%$.
\begin{figure}
    \centering
    \includegraphics[width=0.49\textwidth]{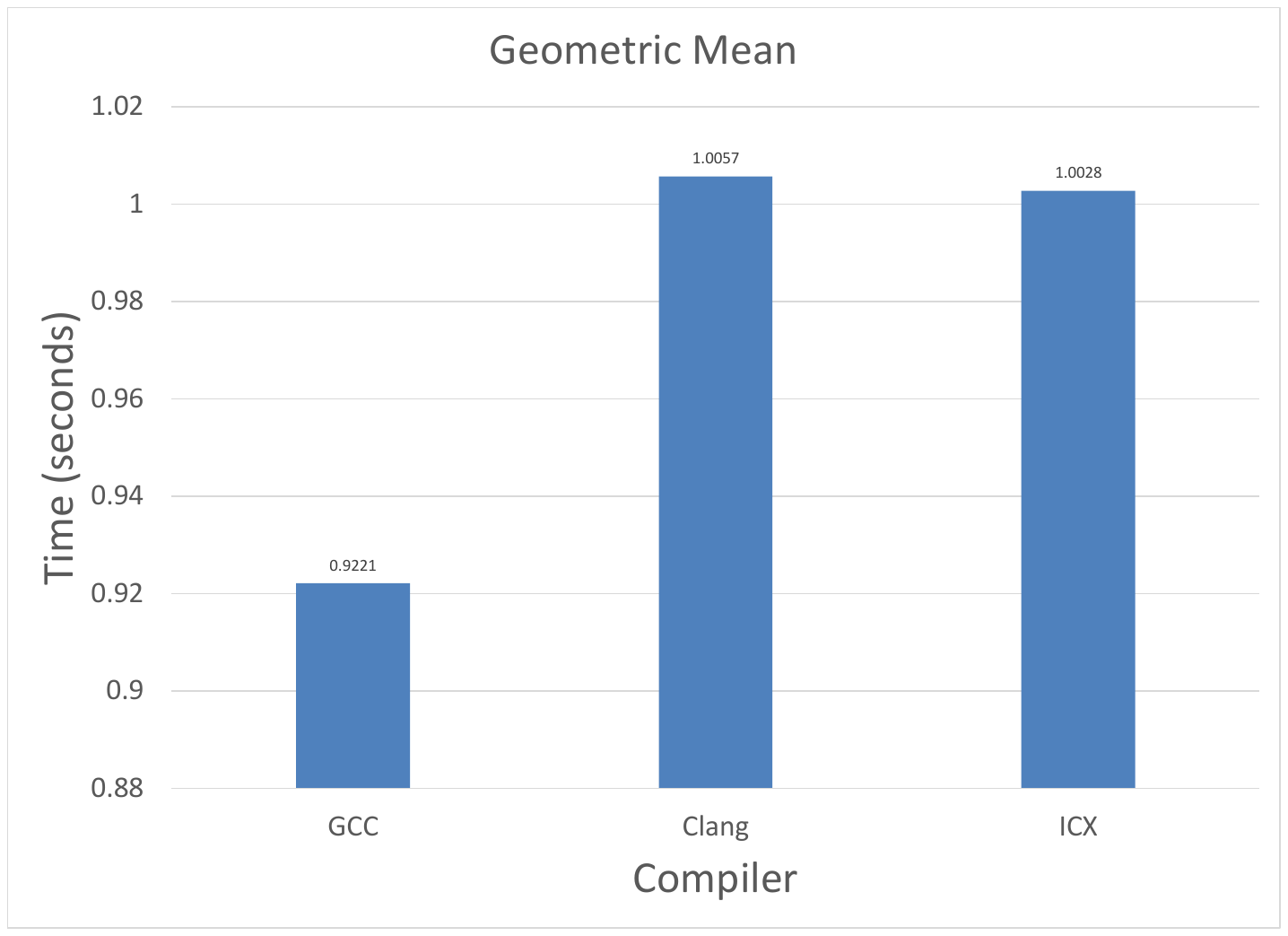}
    \vspace*{-12mm}
	\caption{Geometric Mean of Execution Time}
	\label{fig:avgx}
\end{figure}
Figure~\ref{fig:avgx} shows the geometric mean of the execution time of code vectorized by all three compilers. The code generated by ICX was fastest for $40\%$ of the loops, GCC was fastest for $39\%$ and Clang was fastest for $21\%$. 
Next, we discuss the relative performance of the compilers in greater detail.
\paragraph{GCC}\label{z}
%\subsection{GCC}

%44\% of the loops fall in the range of 0.9 and 1.1.%

\begin{figure}
    \centering
    \includegraphics[width=0.49\textwidth]{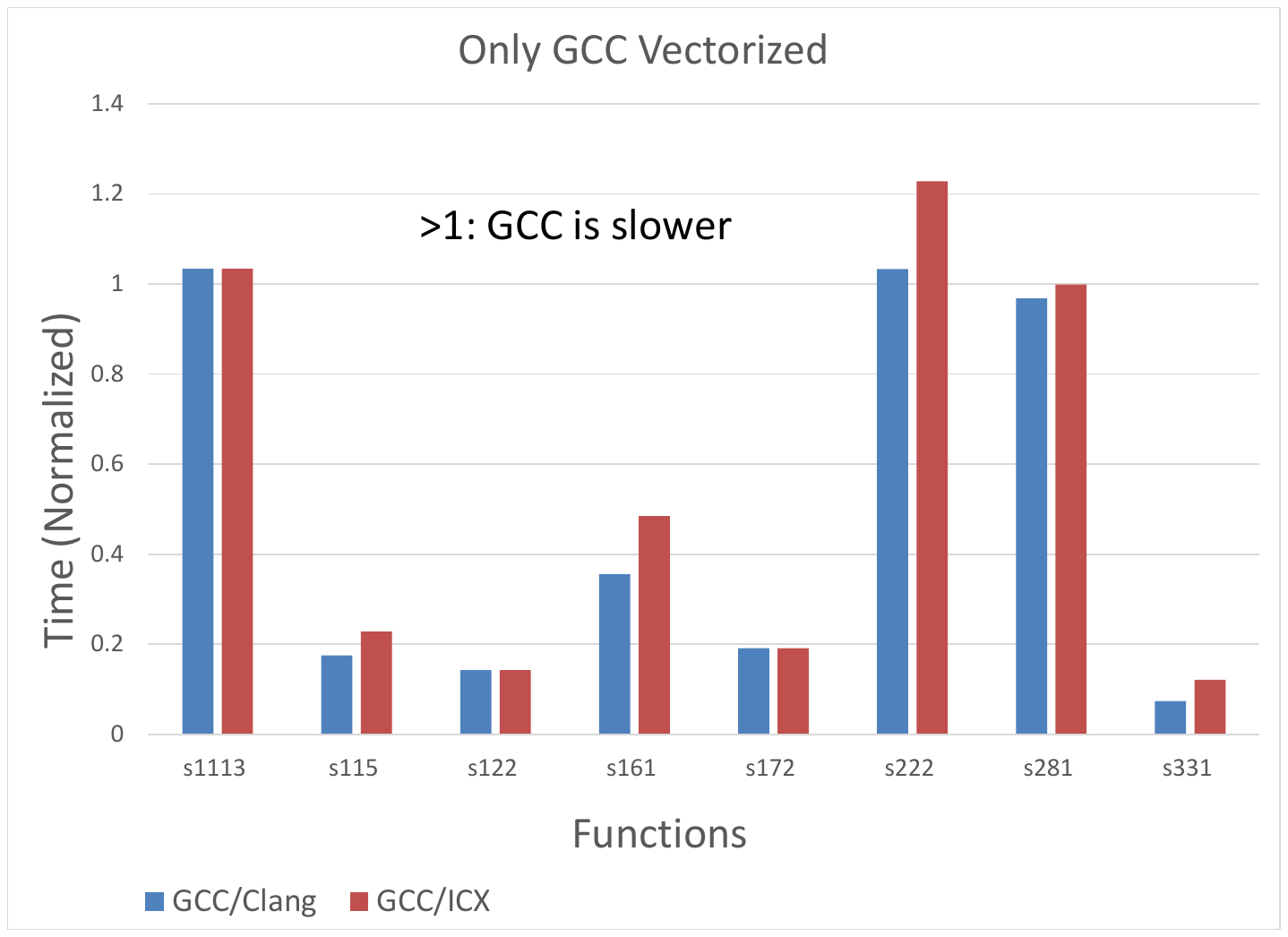}
    \vspace*{-12mm}
	\caption{Execution time of loops vectorized by GCC only}
	\label{fig:onlyg86}
\vspace*{-1mm}
\end{figure}

Figure~\ref{fig:onlyg86} shows the relative execution time of loops reported as having been vectorized by GCC but not Clang or ICX. Out of those 8 loops, GCC was slower than Clang and ICX in 2 cases. One of these loops had a loop carried read-after-write (RAW) dependence which GCC partially vectorized. This optimization did not prove to be beneficial.
Some characteristics seen in the 6 loops where the vectorized code generated by GCC was better include:
\begin{itemize}
    \item Conditional branching
    \item Non-unit but constant stride memory access
    \item Reverse array access
\end{itemize}
\begin{figure}
    \centering
\begin{lstlisting}[style=c]
for (int i = 0; i < lEN_1D; i++)
  x = a[lEN_1D-i-1] + b[i] * c[i];
  a[i] = x-(real_t)1.0;
  b[i] = x;
\end{lstlisting}
    \vspace*{-4mm}
    \caption{Loop \code{s281}}
    \label{fig:s281c}
    \vspace*{-4mm}
\end{figure}
  
%\begin{figure}
   % \centering
    %\lstinputlisting[style=asm]{s281_gcc.asm}
    %\vspace*{-4mm}
    %\caption{Assembly from GCC for loop %\code{s281}}
 %   \label{fig:s281asm}
%\vspace{-6mm}
%\end{figure}

One such loop is \code{s281} as shown in Figure~\ref{fig:s281c}. GCC used \asm{neg} instruction to ensure reverse order access of \code{a[]} on line 2.
%and Figure~\ref{fig:s281asm} shows the assembly %of the inner loop of function s281 by GCC.
%\reg{\%rax} is the index and the \asm{neg} in line 3 ensures \reg{\%rdx}
%becomes the index for reverse order access of %\code{a[]}in line 4.
\begin{figure}
    \centering
\includegraphics[width=0.49\textwidth]{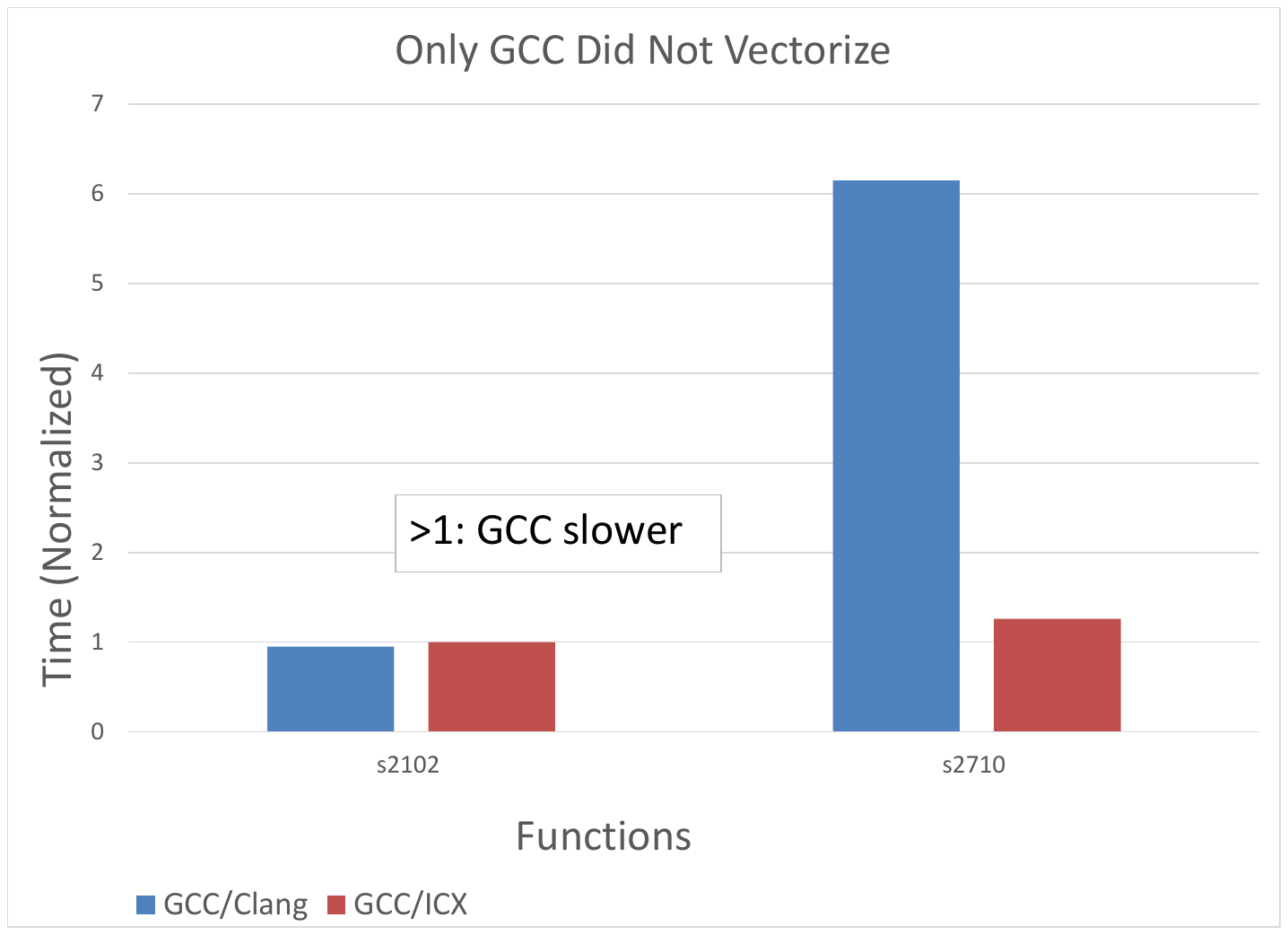}
    \vspace*{-12mm}
	\caption{Execution time of loops not vectorized by GCC only}
	\label{fig:onlynotg86}
 \vspace*{-6mm}
\end{figure}
Figure~\ref{fig:onlynotg86} shows the relative execution time of loops reported as having been vectorized by both Clang and ICX, but not GCC. 
Loop \code{s2102} creates an identity matrix by setting the diagonal elements to one and everything else to zero. Both Clang and ICX used vector scatter instructions which did not provide any performance improvement over non-vectorized stores.  
\paragraph{Clang}

\begin{figure}
    \centering
    \includegraphics[width=0.49\textwidth]{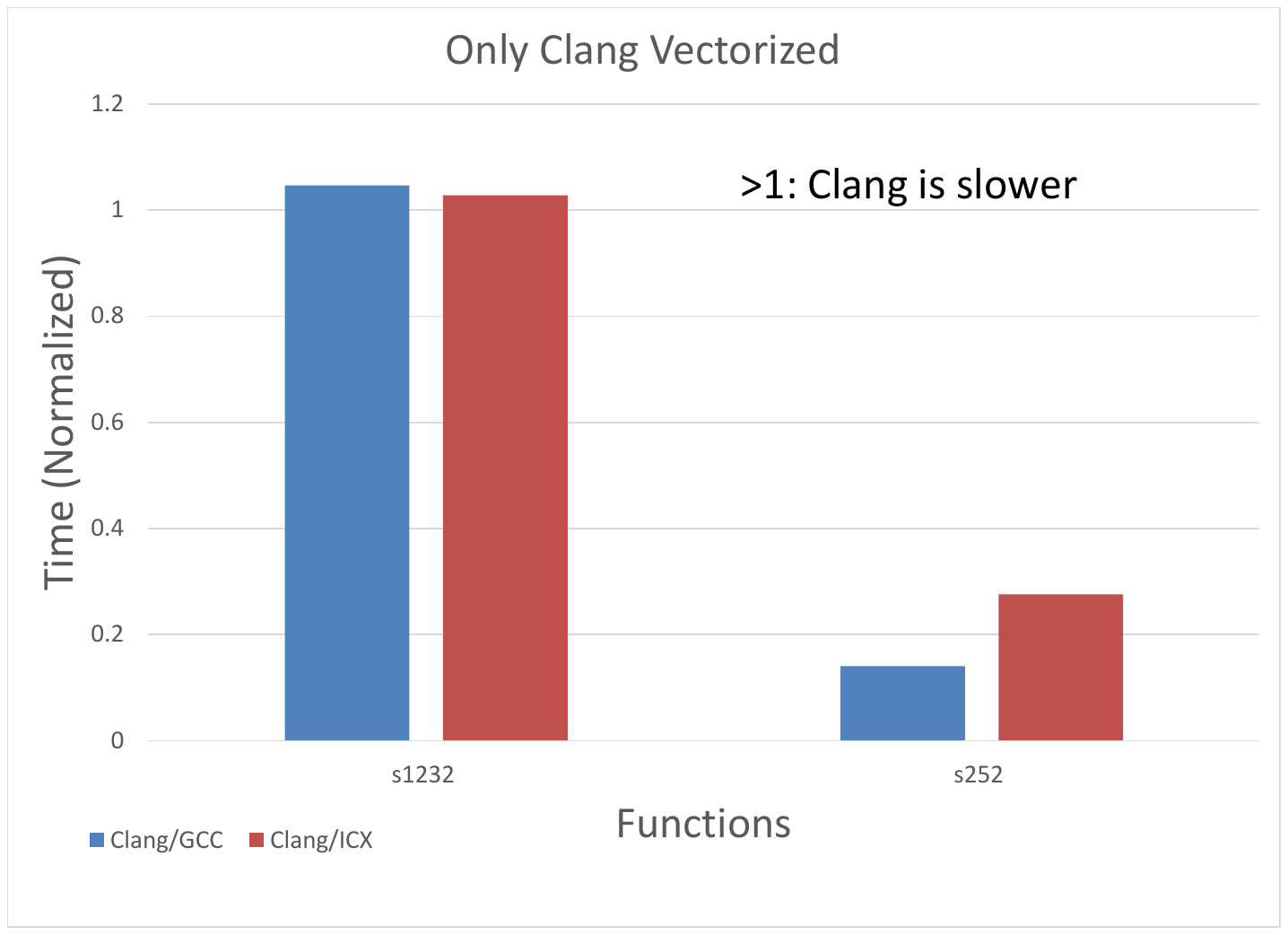}
    \vspace*{-12mm}
    \caption{Execution time of loops vectorized by Clang only}
    \label{fig:onlyc86}
    \vspace*{-2mm}
\end{figure}

Figure~\ref{fig:onlyc86} shows two loops that were vectorized by Clang only.
%Vectorized loop \code{s252} was slower compared %to non-vectorized code produced by GCC and ICX.%
\begin{figure}
    \centering
\begin{lstlisting}[style=c]
for (int j = 0; j < lEN_2D; j++) 
  for (int i = j; i < lEN_2D; i++) 
    (*aa)[i][j] = (*bb)[i][j] + (*cc)[i][j];
       
\end{lstlisting}
    \vspace*{-2mm}
    \caption{Loop \code{s1232}}
    \label{fig:s1232c}
\vspace*{-2mm}
\end{figure}  
\begin{figure}
    \centering
\begin{lstlisting}[style=asm]
vpmullq     %zmm8,%zmm0,%zmm1
kxnorw      %k0,%k0,%k1
vxorps      %xmm2,%xmm2,%xmm2
kxnorw      %k0,%k0,%k2
vgatherqps  (%r10,%zmm1,4),%ymm2{%k1}
vxorps      %xmm3,%xmm3,%xmm3
vgatherqps  (%r11,%zmm1,4),%ymm3{%k2}
vaddps      %ymm3,%ymm2,%ymm2
kxnorw      %k0,%k0,%k1
vscatterqps %ymm2,(%rbx,%zmm1,4){%k1}
vpaddq      %zmm10,%zmm0,%zmm0
add         $0x8,%r9
jne         30000 <s1232+0x4e0>
\end{lstlisting}
    \vspace*{-2mm}
    \caption{Assembly from Clang for loop \code{s1232}}
    \label{fig:s1232asm}
    \vspace*{-2mm}
\end{figure}

Figure~\ref{fig:s1232c} shows the C code for loop \code{s1232}. The stride for all 2D arrays are constant (the size of row). Figure~\ref{fig:s1232asm} is the assembly produced by Clang. The loads and stores are performed using masked gather and scatter instructions, which did not yield any improvement over the unvectorized code produced by GCC and ICX.
\begin{figure}
    \centering
    \includegraphics[width=0.49\textwidth]{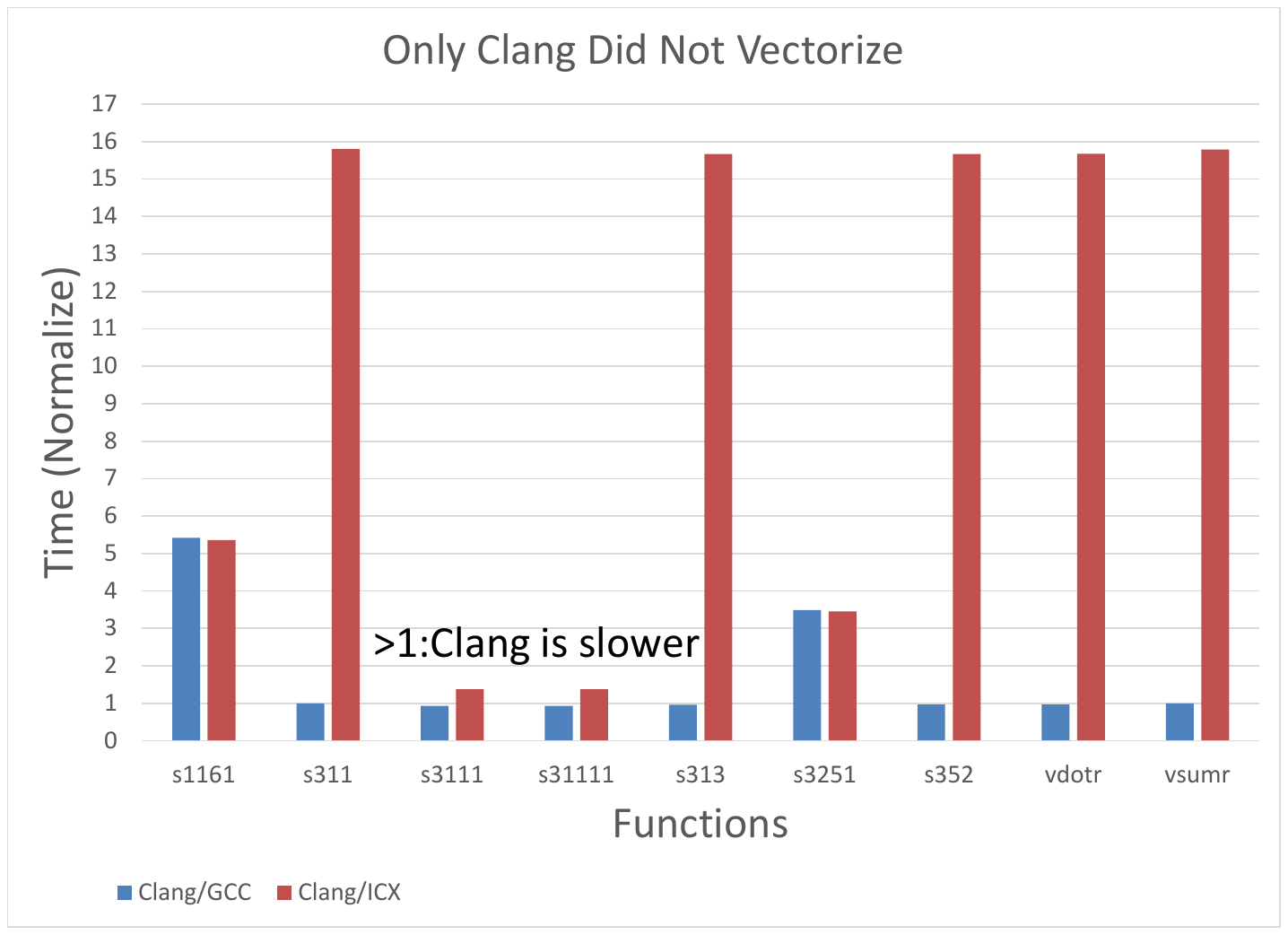}
    \vspace*{-12mm}
	\caption{Execution time of loops not vectorized by Clang only}
	\label{fig:onlynotc86}
 \vspace*{-1mm}
\end{figure}
Figure~\ref{fig:onlynotc86} shows the normalized execution time of loops not vectorized by Clang but vectorized by both GCC and ICX. 7 of these loops perform reductions. The vectorized code generated by GCC was not noticeably faster than the unvectorized code generated by clang. We examined the vectorized code generated by ICX and GCC for one such loop, \code{s3111}.

\begin{figure}
    \centering
\begin{lstlisting}[style=c]
for (int i = 0; i < lEN_1D; i++)
  if (a[i] > (real_t)0.)
    sum += a[i]        
\end{lstlisting}
    \vspace*{-4mm}
    \caption{Loop \code{s3111}}
    \label{fig:s3111c}
    \vspace*{-3mm}
\end{figure} 

Figure~\ref{fig:s3111c} is the C code for loop \code{s3111}. GCC partially vectorized this using vector load instruction. ICX, on the other hand, used vector loads, compares and adds. The temporary store also used vector instructions. 
\paragraph{ICX}
\begin{figure}
    \centering
    \includegraphics[width=0.49\textwidth]{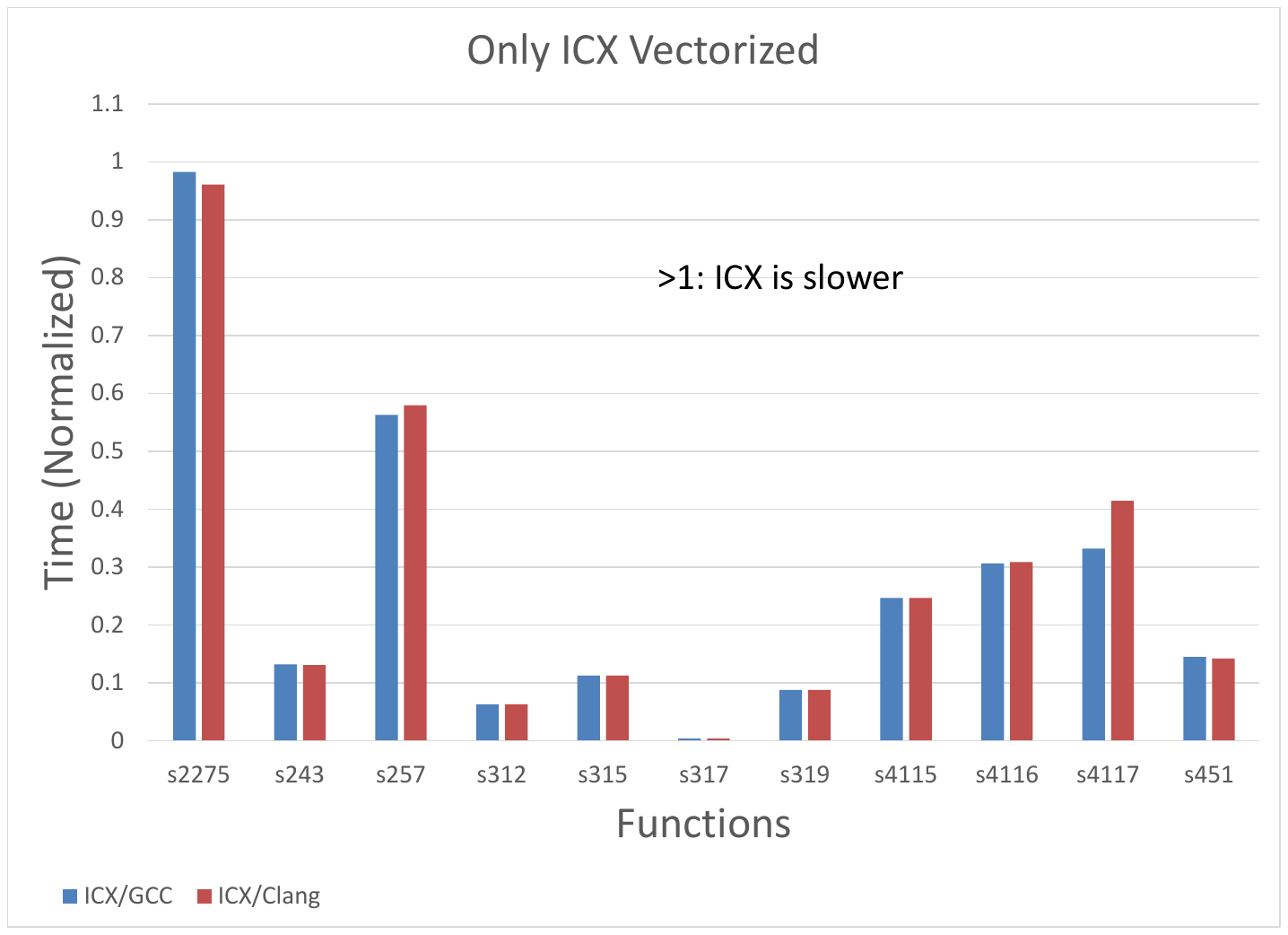}
    \vspace*{-12mm}
	\caption{Execution time of loops vectorized by ICX only}
	\label{fig:onlyi86}
 \vspace*{-4mm}
\end{figure}
Figure~\ref{fig:onlyi86} shows the loops vectorized by ICX, but not the other compilers. Some common features shared by these loops are:
\begin{itemize}
    \item Indirect addressing
    \item Reductions
    \item Non-unit stride access
\end{itemize}
\begin{figure}
    \centering
\begin{lstlisting}[style=c]
q = (real_t)1.
  for (int i = 0; i < lEN_1D/2; i++) 
    q *= (real_t).99
        
\end{lstlisting}
    \vspace*{-4mm}
    \caption{Loop \code{s317}}
    \label{fig:s317c}
    \vspace*{-4mm}
\end{figure}

Figure~\ref{fig:s317c} is the C code for loop \code{s317}. The statement in line 3 can be re-written in closed form as \code{\footnotesize $q=0.99^n$} where n is the loop trip count.
\begin{figure}
    \centering
\begin{lstlisting}[style=asm]
vmulps %zmm1,%zmm0,%zmm0 //%zmm0=q, %zmm1=0.99
add           $0x10,%eax
cmp           %r12d,%eax
jl            4468a0 <s317+0x1a0>
vextractf64x4 $0x1,%zmm0,%ymm1
vmulps        %zmm1,%zmm0,%zmm0
vextractf128  $0x1,%ymm0,%xmm1
vmulps        %xmm1,%xmm0,%xmm0
vshufpd       $0x1,%xmm0,%xmm0,%xmm1
vmulps        %xmm1,%xmm0,%xmm0
vmovshdup     %xmm0,%xmm1
vmulss        %xmm1,%xmm0,%xmm0
\end{lstlisting}
    \vspace*{-4mm}
    \caption{Assembly from ICX for loop \code{s317}}
    \label{fig:s317asm}
\vspace*{-4mm}
\end{figure}
Figure~\ref{fig:s317asm} shows part of the assembly generated by ICX. Lines 1 to 4 perform the multiplication in line 3 of Figure~\ref{fig:s317c}. The initial value of \reg{\%zmm0} is set to 1 (not shown in the listing). Using vector multiplication, the number of iterations is reduced from \code{len2D/2} to \code{(len2D/2)/16}. The individual values in \reg{\%zmm0} are multiplied Lines in 5 to 12 which results in the closed form solution.
\begin{figure}
    \centering
    \includegraphics[width=0.49\textwidth]{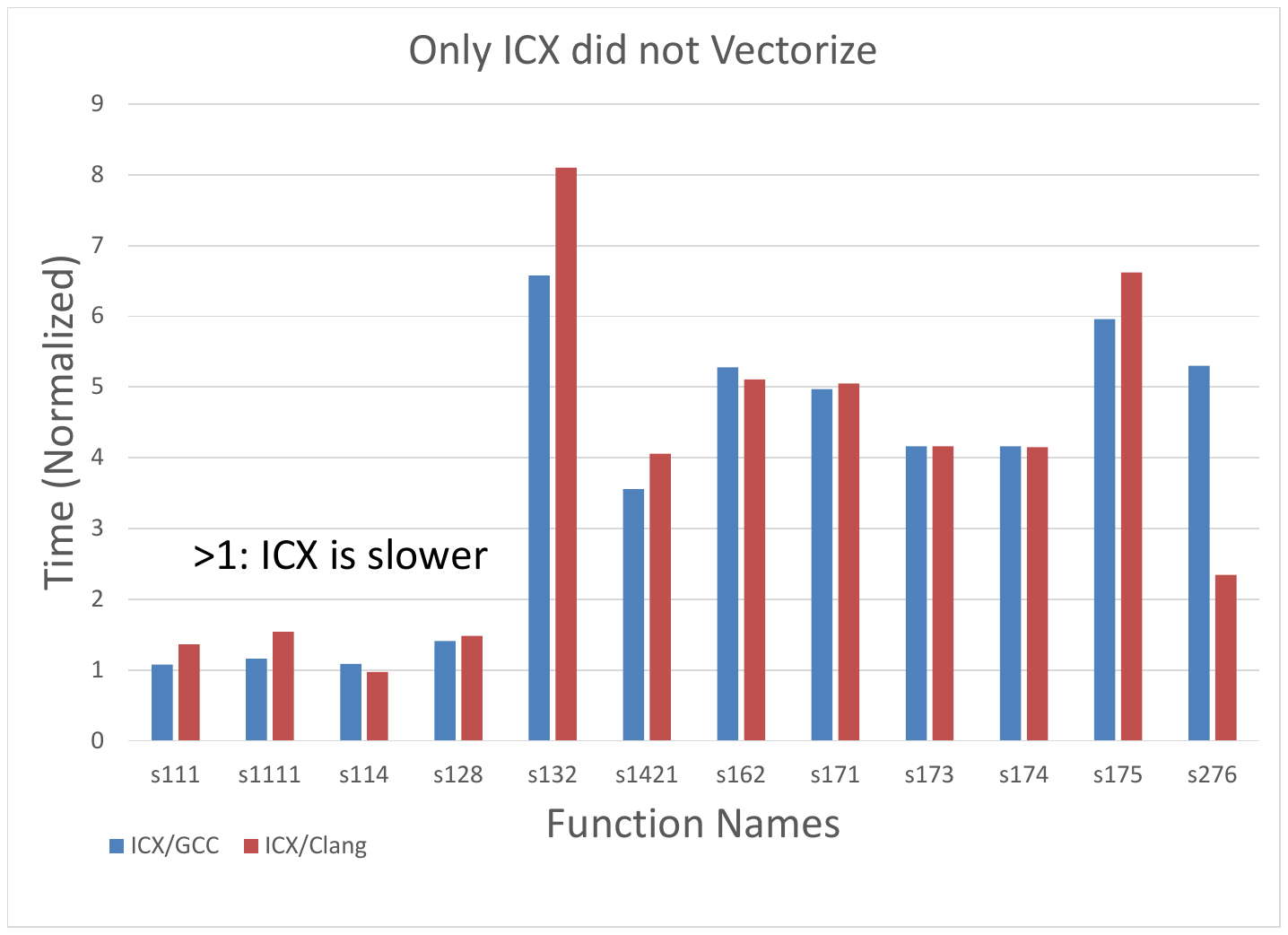}
    \vspace*{-12mm}
	\caption{Execution time of loops not vectorized by ICX only}
	\label{fig:onlyin86}
\end{figure}

Figure~\ref{fig:onlyin86} shows the normalized execution time of loops not vectorized by ICX only. Nearly 50\% of these loops have array subscripts that are linear functions of the loop iterator and some induction variable other than the loop iterator.

\subsection{ARM}
We used the following flags when compiling the test-suite on ARM: \code{\footnotesize-O3 -mcpu=a64fx+sve -msve-vector-bits=512}. The vectorization reports were generated using the same options as x86.
%\vspace{-2mm}
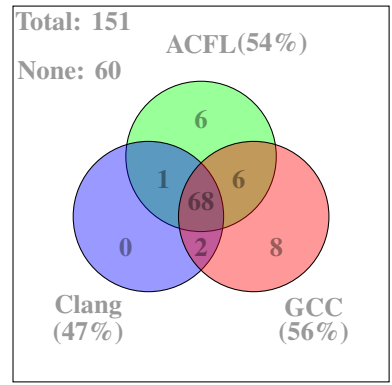
\begin{figure}[t]
\centering
\begin{tikzpicture}
%% You can adjust the opacity here. For venn diagrams, it is convenient to have a low opacity so that you can see intersections
	\begin{scope} [fill opacity = .4]
%% The draw command knows a lot of shapes. To make a rectangle, you need to specify two diagonal corners. Ensure you always have a semicolon at the end of your draw commands, otherwise latex flips out.
    %\draw (-2,2.5) rectangle (2,-0.5);
    \draw (0,0) rectangle (5,5);
%% Similarly, you can make a circle by specifying the center and then the radius. You can also add a fill color, but if you're printing in black and white you'll probably want to remove that line.
    \draw[fill=green, draw = black] (2.5,3) circle (1);
    \draw[fill=blue, draw = black] (1.8,2.2) circle (1);
    \draw[fill=red, draw = black] (3.2,2.2) circle (1);
    %\draw[fill=red, draw = black] (0,-2) circle (3);
%% We can use the node command to label points. If you put your cursor on "LARGE" or "textbf" a box will drop down with size and text style options.
    %\node at (-2,3.2) {\LARGE\textbf{X}};
    \node at (2.5,4.5) {\color{black}\textbf{ACFL}};
        \node at (3.45,4.5) {\textbf{\color{black}(54\%)}};
    \node at (4,1) {\color{black}\textbf{GCC}};
        \node at (4,0.65) {\color{black}\textbf{(56\%)}};
    \node at (1,1) {\color{black}\textbf{Clang}};
    \node at (1,0.65) {\color{black}\textbf{(47\%)}};
    \node at (2.5,2.4) {\textbf{68}}; %all y%
    \node at (2.5,1.8) {\textbf{2}}; %gycyin%
    \node at (1.5,1.8) {\textbf{0}}; %cy%
     \node at (2,2.7) {\textbf{1}}; %gncyiy%
    \node at (2.5,3.5) {\textbf{6}}; %iy%
    \node at (3,2.7) {\textbf{6}}; %gycniy y%
    \node at (3.5,1.8) {\textbf{8}}; %gy%
    \node at (0.8,4.8) {\textbf{Total: 151}};
    \node at (0.75,4.15) {\textbf{None: 60}};

    \end{scope}
%% And now you have a venn diagram. Yay!
%\draw[help lines](-5,5) grid (5,-6);    This line can draw the grid lines to help guide you. I use these when I'm writing the code and then delete this line when I publish the pdf.
\end{tikzpicture}
    \caption{Loops vectorized by GCC, ACFL, and Clang on ARM}
    \label{fig:venarm}
\vspace{-4mm}    
\end{figure}
Figure~\ref{fig:venarm} shows the number of loops vectorized by GCC, Clang, and ACFL. Out of the 151 loops, GCC did not vectorize $44\%$, Clang did not vectorize $53\%$ and ACFL did not vectorize $46\%$. These results are similar to those reported by Bine~\textit{et al.}~\cite{new:pa}.
%\vspace{-2mm} 
\begin{figure}
    \centering
    \includegraphics[width=0.49\textwidth]{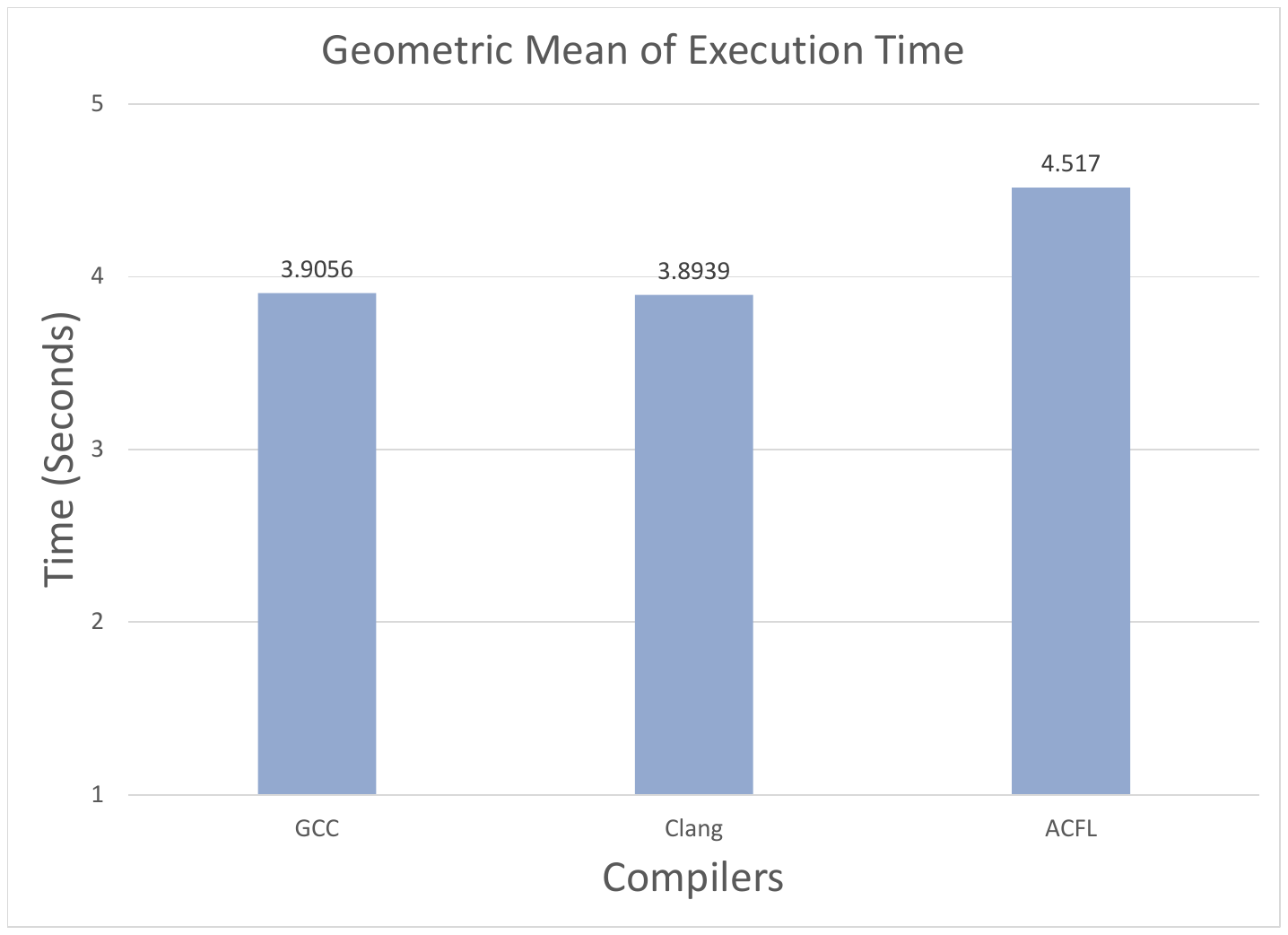}
    \vspace*{-12mm}
    \caption{Geometric Mean of Execution Time}
    \label{fig:avgarm}
    \vspace*{-4mm}
\end{figure}

Figure~\ref{fig:avgarm} shows the geometric mean of the execution time of code vectorized by all three compilers. Of these, the code generated by Clang was fastest for $65\%$ of the loops, GCC was fastest for $22\%$ and ACFL was fastest for $13\%$.

\paragraph{GCC}
\begin{figure}
    \centering
    \includegraphics[width=0.49\textwidth]{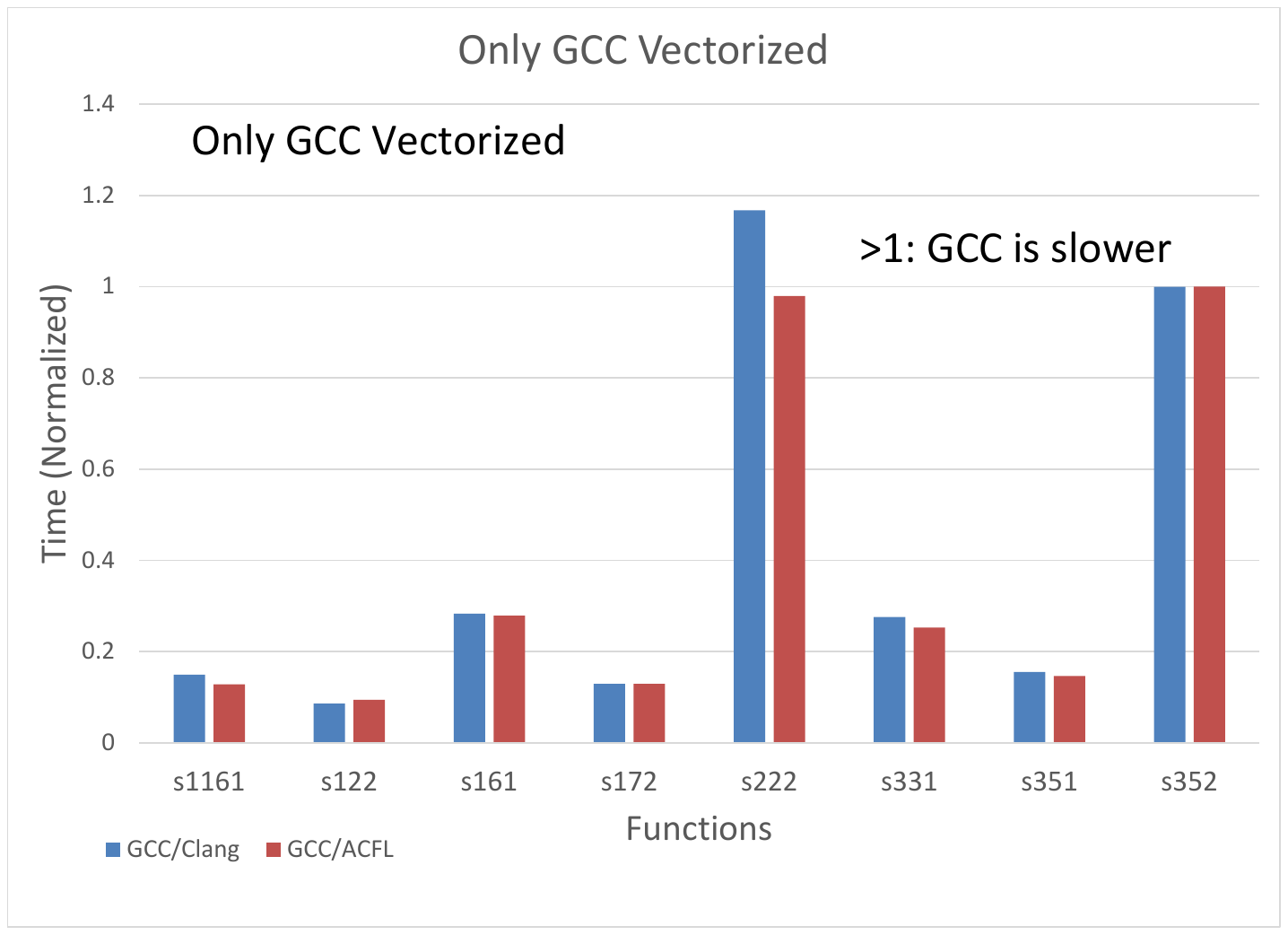}
    \vspace*{-12mm}
	\caption{Execution time of loops vectorized vectorized by GCC only}
	\label{fig:onlygARM}
 \vspace*{-2mm}
\end{figure}
Figure~\ref{fig:onlygARM} shows the normalized execution time of loops that were vectorized by GCC but not by either Clang or ACFL. \code{s2710} is the only loop that was vectorized by both ACFL and Clang, but not GCC.
Most of these loops are the same as those in Figure~\ref{fig:onlyg86}. 
\begin{figure}
    \centering
\begin{lstlisting}[style=c]
for (int i = 1; i < lEN_1D; i++)
  a[i] += b[i] * c[i]
  e[i] = e[i - 1] * e[i - 1]
  a[i] -= b[i] * c[i]
        
\end{lstlisting}
    \vspace*{-4mm}
    \caption{Loop \code{s222}}
    \label{fig:s222_arm}
\end{figure}

Despite being vectorized, the performance of loops \code{s222} and \code{s352} did not improve. Figure~\ref{fig:s222_arm} is the C code for loop \code{s222}. GCC vectorized only the statements in line 2 and 4. The statement in line 3 has a read-after-write (RAW) dependency which cannot be vectorized. 
\begin{figure}
    \centering
\begin{lstlisting}[style=c]
for (int i = 0; i < lEN_1D; i += 5) 
  dot = dot + a[i] * b[i] +
       a[i + 1] * b[i + 1] +
       a[i + 2]* b[i + 2] +
       a[i + 3] * b[i + 3] +
       a[i + 4] * b[i + 4]
\end{lstlisting}
    \vspace*{-4mm}
    \caption{Loop \code{s352}}
    \label{fig:s352_arm}
\end{figure}

Figure~\ref{fig:s352_arm} is the C code for loop \code{s352}. Even though Clang and ACFL did not vectorize it, both unrolled it by factors of 20 and 24 respectively, which might have resulted in the speedup.
\paragraph{Clang}
\begin{figure}
    \centering
    \includegraphics[width=0.49\textwidth]{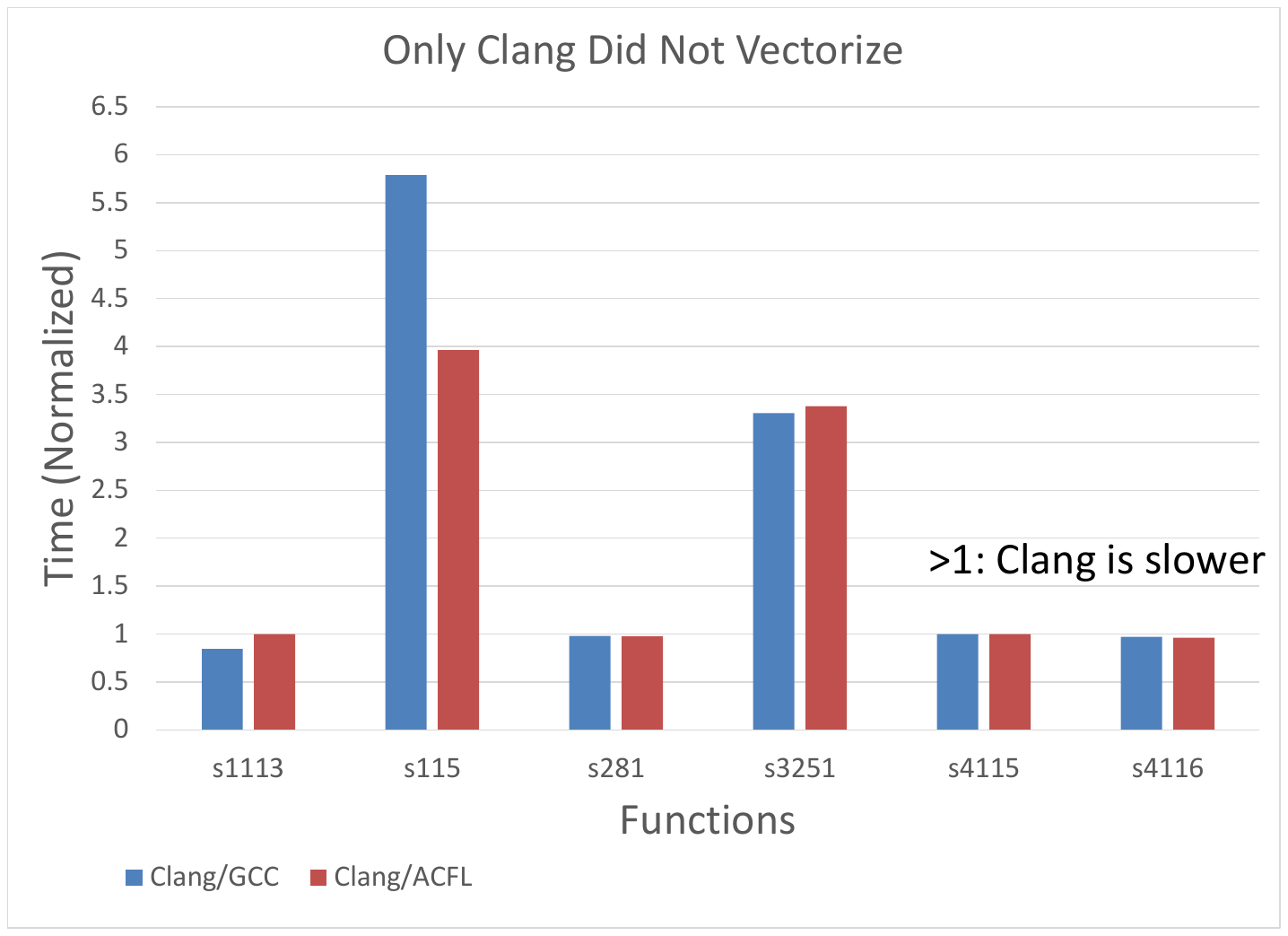}
    \vspace*{-12mm}
	\caption{Execution time of loops not vectorized by Clang only}
	\label{fig:onlycARM}
 \vspace*{-2mm}
\end{figure}
Figure~\ref{fig:onlycARM} shows the normalized execution time of loops that were not vectorized by Clang, but were vectorized by the other compilers. Every loop vectorized by Clang was vectorized by either ACFL or GCC. Some of the features of the 4 loops where Clang was not slower despite not vectorizing are:
\begin{itemize}
    \item Write-after-read dependence (WAR) for a single iteration
    \item Reverse data access
    \item Indirect memory lookup
\end{itemize}
\begin{figure}
    \centering
\begin{lstlisting}[style=c]
for (int i = 0; i < lEN_1D; i++)
  sum += a[i] * b[ip[i]]
\end{lstlisting}
    \vspace*{-2mm}
    \caption{Loop \code{s4115}}
    \label{fig:s4115_arm}
    \vspace*{-2mm}
\end{figure}

Figure~\ref{fig:s4115_arm} is the C code for loop \code{s4115}.  
\begin{figure}
    \centering
\begin{lstlisting}[style=asm]
ld1w    {z1.s}, p0/z, [x19, x8, lsl #2]
ld1w    {z0.s}, p0/z, [x20, x8, lsl #2]
add     x8, x8, x24
ld1w    {z1.s}, p0/z, [x21, z1.s, sxtw #2]
fmul    z0.s, z0.s, z1.s
fadda   s2, p0, s2, z0.s
whilelo	p0.s, x8, x22
b.mi	43466c <s4115+0xa4>
\end{lstlisting}
    \vspace*{-2mm}
    \caption{Assembly from ACFL for loop \code{s4115}}
    \label{fig:s4115ARM}
\end{figure}
Figure~\ref{fig:s4115ARM} is the assembly generated by ACFL. GCC produced similar assembly. The result of the load instruction in line 1 is stored in \reg{z1} which is used in the offset calculation for the gather instruction on line 4. 
\begin{figure}
    \centering
\begin{lstlisting}[style=asm]
mov	x8, xzr
ldpsw	x11, x12, [x10, #-16]
ldp	s1, s2, [x9, #-16]
sub	x8, x8, #0x8
ldr	s0, [x21, x11, lsl #2]
fmadd	s0, s1, s0, s8
\end{lstlisting}
    \vspace*{-2mm}
    \caption{Assembly from Clang for loop \code{s4115}}
    \label{fig:s4115_cl_ARM}
\end{figure}

Clang, on the other hand, did not vectorize the loop, but unrolled it by a factor of 8 instead. Also, the \asm{ldpsw} instruction was used which loads a pair of words. The first set of unrolled instructions is presented in Figure~\ref{fig:s4115_cl_ARM}. It is not clear why the vector gather instructions were not profitable. 
%did not result in faster execution time because it had to wait till all of the index values (for %one vector register) arrive from memory.

\begin{figure}
    \centering
    \includegraphics[width=0.49\textwidth]{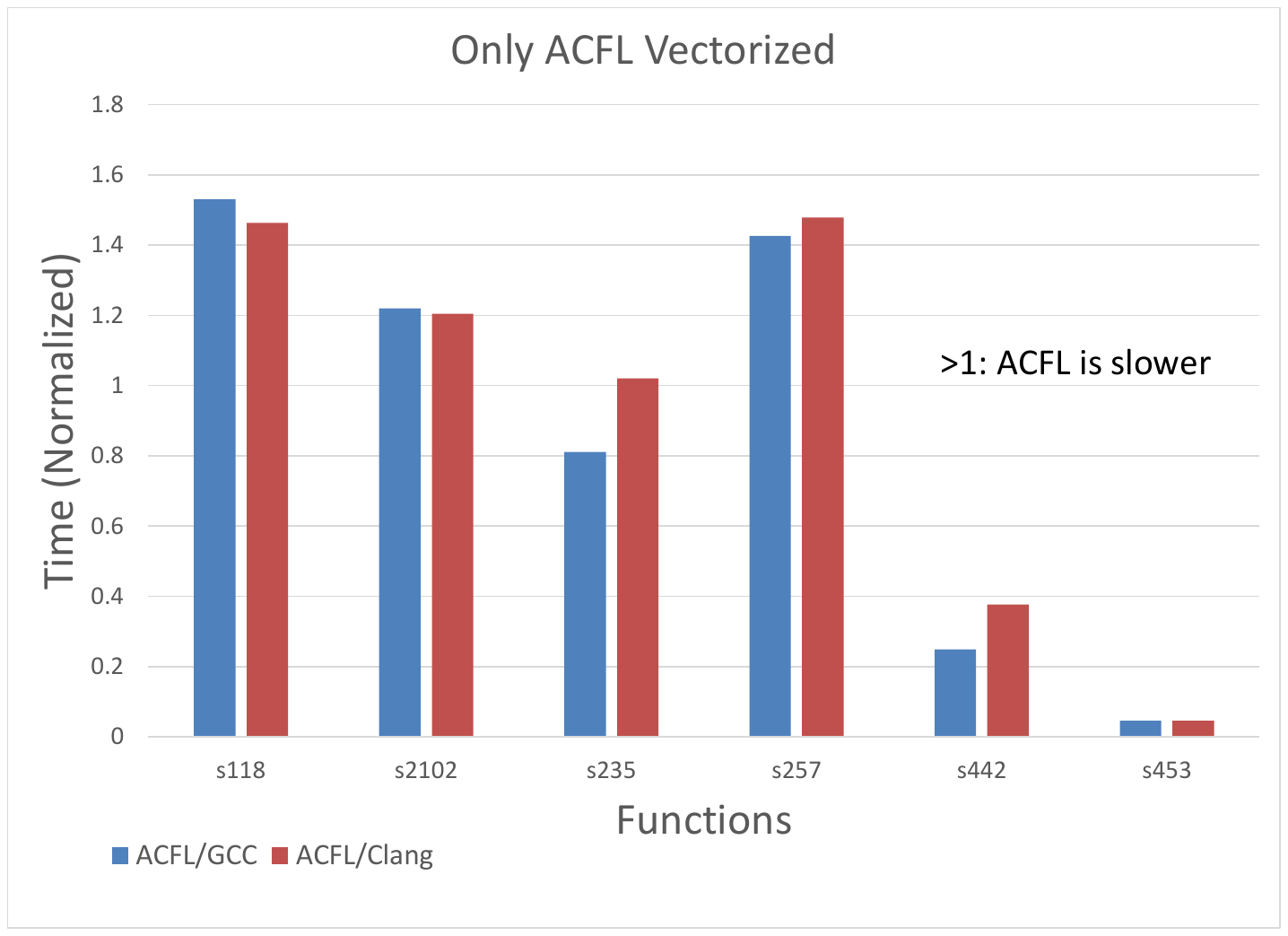}
    \vspace*{-12mm}
	\caption{Execution time of loops vectorized  by ACFL only}
	\label{fig:onlyacflARM}
 \vspace*{-6mm}
\end{figure}

\begin{figure}
    \centering
    \includegraphics[width=0.49\textwidth]{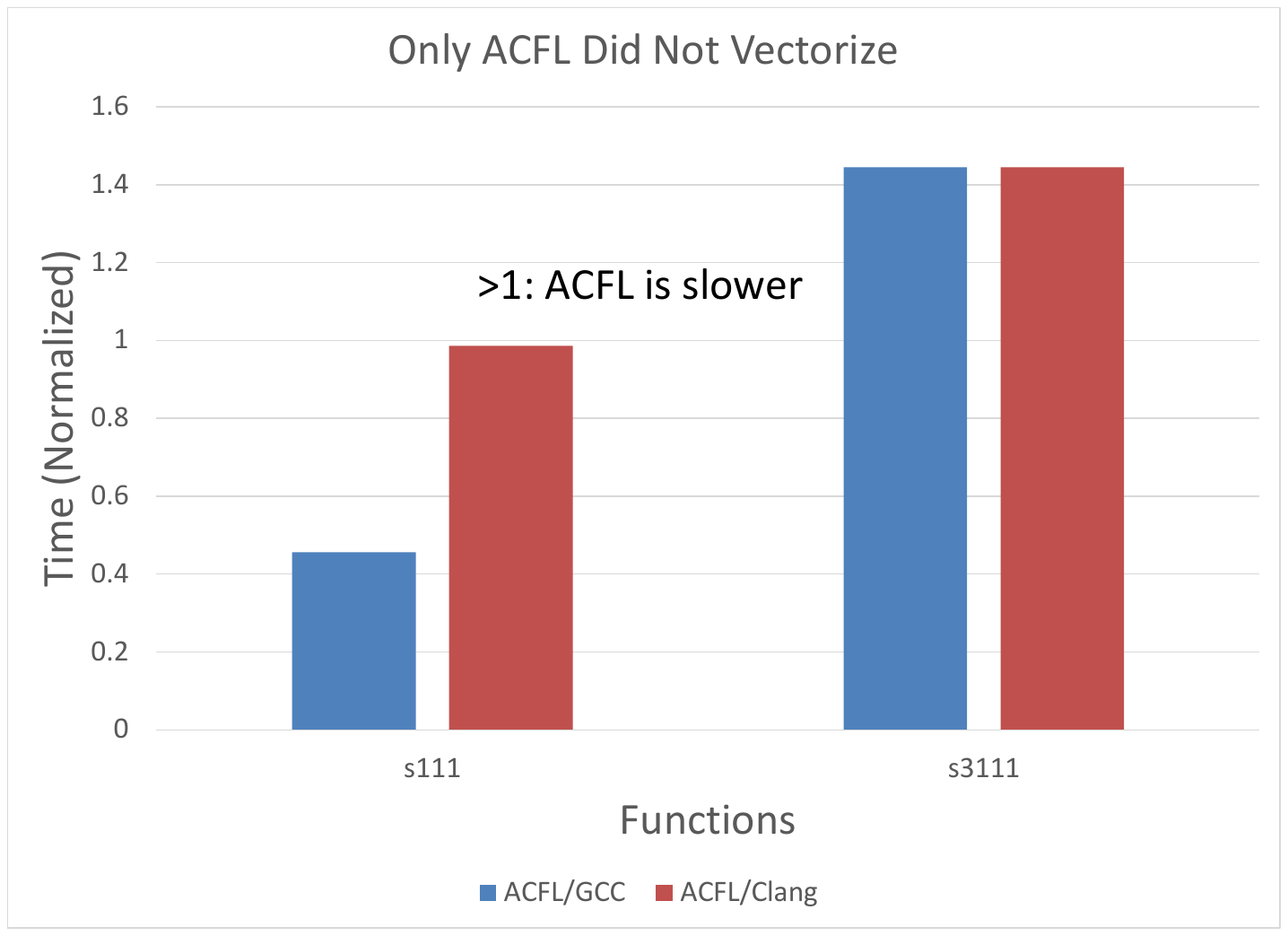}
    \vspace*{-12mm}
	\caption{Execution time of loops not vectorized  by ACFL only}
	\label{fig:onlyacflARM_n}
 \vspace*{-4mm}
\end{figure}

\paragraph{ACFL}
Figure~\ref{fig:onlyacflARM} shows 4 loops that were vectorized by ACFL, but not by any of the other compilers. Figure~\ref{fig:onlyacflARM_n} shows 2 loops that were vectorized by both GCC and Clang, but not ACFL.


\begin{figure}
    \centering
\begin{lstlisting}[style=c]
for (int i = 0; i < lEN_1D; i++)
  s += (real_t)2.;
  a[i] = s * b[i];
        
\end{lstlisting}
    \vspace*{-3mm}
    \caption{Loop \code{s453}}
    \label{fig:s453_acfl}
\vspace*{-3mm}
\end{figure}
\begin{figure}
    \vspace{-2mm}
    \centering
\begin{lstlisting}[style=asm]
scvtf   s0, x8           
ld1w    {z1.s}, p0/z, [x21, x8, lsl #2]
fadd    s0, s0, s0 
mov     z0.s, s0
fadd    z0.s, z0.s, z2.s
fadd    z0.s, z0.s, z3.s 
fmul    z0.s, z0.s, z1.s
st1w    {z0.s}, p0, [x20, x8, lsl #2]
add     x8, x8, x24
whilelo p0.s, x8, x28
\end{lstlisting}
    \vspace*{-3mm}
    \caption{Assembly from ACFL for loop \code{s453}}
    \label{fig:s453_acfl_asm}
\vspace*{-3mm}
\end{figure}

Only two vectorized loops were faster. Figure~\ref{fig:s453_acfl} is the C code for loop \code{s453}. The computations in line 2 and 3 can be re-written as \code{\footnotesize a[i]=(2*i+2)*b[i]}, which is what ACFL did as shown in Figure~\ref{fig:s453_acfl_asm}.
Another loop that was vectorized only by ACFL and showed better performance is loop \code{s442}, which contains a \code{switch} statement. ACFL was able to vectorize it using \asm{cmpeq} and \asm{sel} instructions and predicate registers which can perform loads and other operations conditionally. 

\subsection{X86 and ARM}
This section summarizes the difference in behavior of GCC and Clang across the two platforms. Almost all loops that were vectorized by GCC or Clang on one platform were vectorized on the other. We found 3 loops that were not vectorized by GCC on X86 which were vectorized on ARM. 2 of them used vector gather instructions. We also found 3 loops vectorized by Clang on x86 but not on ARM. 2 of these had non-unit but constant stride memory accesses. 4 loops were reported vectorized by Clang on ARM but not on x86, all of which performed reductions. 

Out of 65 loops that were vectorized by both GCC and Clang on x86, GCC outperformed Clang in 34 ($52\%$) of the cases. For ARM, out of 70 loops vectorized by both, Clang outperformed GCC in 51 ($73\%$) of the cases. Of the loops that were vectorized by both GCC and Clang on both x86 and ARM, the code produced by GCC outperformed that produced by Clang in 14 cases, while Clang outperformed GCC in 26 cases.

\subsection{Indirect Memory Access}
There are 8 loops in TSVC2 with indirect memory accesses. Both x86 and ARM provide vector gather/scatter instructions. Neither GCC nor Clang were able to utilize them to vectorize these 8 loops on x86. However, ICX was able to vectorize 2 loops. On ARM, GCC and ACFL were able to vectorize the same 2 loops but Clang was not able to vectorize any. 
We discussed \code{s4115} in Figure~\ref{fig:s4115_arm}, which was vectorized on ARM by GCC and ACFL but not Clang. GCC was not able to vectorize it on x86. 
\begin{figure}
    \centering
\begin{lstlisting}[style=c]
for (int i = 0; i < lEN_1D; i++)
  a[i] += b[ip[i]] * s;
        
\end{lstlisting}
    \vspace*{-4mm}
    \caption{Loop \code{s4112}}
    \label{fig:s4112}
\end{figure}
Figure~\ref{fig:s4112} is the C code for loop \code{s4112}. Despite being similar to loop \code{s4115}, no compiler, on either platform, vectorized it.

\section{Conclusion and Future Work}
We have investigated the ability of several compilers to vectorize on two different hardware platforms. $35\%$ of the loops in the TSVC2 suite were vectorized by all three compilers on x86 and $36\%$ were not vectorized by any of them. For ARM, these numbers are $45\%$ and $40\%$ respectively. GCC reported more loops being vectorized than Clang on both X86 and ARM. However, of the loops vectorized by both GCC and Clang, the code generated by GCC performed better on x86 whereas code generated by Clang performed better on ARM. There were cases where the compilers would vectorize a loop on X86 but not on ARM (and vice versa). There were no immediately obvious, consistent strengths or weaknesses in any one compiler's ability to vectorize. It was also unclear when the code generated by any given compiler would outperform the others. We have reported the few patterns that were apparent to us. In future work, we intend to focus on loops from specific domains, which could help determine if any compiler is particularly suitable for a given domain.

\section{Acknowledgment}
We would like to thank the anonymous reviewers for their feedback. This work is partially supported by Triad National Security, LLC subcontracts \#581326 and \#C4975. Also, this work was partially supported by The Regents of the University of 
California (Lawrence Berkeley National Laboratory) under Subcontract \#7558382. Any opinions, findings, or conclusions expressed in this paper do not necessarily represent the views of the DOE or the US Government.

\balance
\bibliographystyle{IEEEtran} 
\bibliography{nazmus}

\end{document}